\documentclass[conference,compsoc]{IEEEtran}

\ifCLASSOPTIONcompsoc
  \usepackage[nocompress]{cite}
\else
  \usepackage{cite}
\fi
\usepackage{url}

\usepackage{amsmath}
\usepackage{tikz}
\usepackage{graphicx}

\usepackage{booktabs}
\usepackage{tabularx}
\usepackage{makecell}
\usepackage{array}
\usepackage{pifont}

\usepackage{enumitem}
\usepackage{float}
\usepackage[section]{placeins}

\usepackage{xcolor}
\usepackage[most]{tcolorbox}

\usepackage{listings}

\usepackage{subcaption}

\newcommand{\cmark}{\ding{51}}

\newcommand{\xmark}{\ding{55}}

\newtcolorbox{rqbox}{
  colback=gray!6,
  colframe=gray!60,
  boxrule=0.5pt,
  arc=3pt,
  left=6pt,
  right=6pt,
  top=6pt,
  bottom=6pt
}

\definecolor{codebg}{RGB}{248,248,248}
\definecolor{codeframe}{RGB}{220,220,220}
\definecolor{codekw}{RGB}{0,92,175}
\definecolor{codestring}{RGB}{163,21,21}
\definecolor{codecomment}{RGB}{0,128,0}

\lstdefinestyle{code}{
    backgroundcolor=\color{codebg},
    frame=single,
    rulecolor=\color{codeframe},
    basicstyle=\ttfamily\footnotesize,
    keywordstyle=\color{codekw}\bfseries,
    stringstyle=\color{codestring},
    commentstyle=\color{codecomment}\itshape,
    breaklines=true,
    breakatwhitespace=true,
    columns=fullflexible,
    keepspaces=true,
    showstringspaces=false,
    captionpos=b
}

\lstdefinelanguage{json}{
morestring=[b]",
morekeywords={true,false,null},
sensitive=false,
morecomment=[l]{//},
morecomment=[s]{/*}{*/}
}

\begin{document}

\date{}
\title{AXE: Grey-Box Exploitability Confirmation for Localized Vulnerability Reports}

\author{
\IEEEauthorblockN{
Amirali Sajadi\IEEEauthorrefmark{1},
Tu Nguyen\IEEEauthorrefmark{1},
Kostadin Damevski\IEEEauthorrefmark{2},
and Preetha Chatterjee\IEEEauthorrefmark{1}
}
\IEEEauthorblockA{\IEEEauthorrefmark{1}Drexel University, Philadelphia, PA, USA\\
\{amirali.sajadi, jn866, preetha.chatterjee\}@drexel.edu}
\IEEEauthorblockA{\IEEEauthorrefmark{2}Virginia Commonwealth University, Richmond, VA, USA\\
kdamevski@vcu.edu}
}

\maketitle

\begin{abstract}
    Vulnerability detection tools are widely adopted in software projects, yet they often overwhelm maintainers with false positives and non-actionable reports. Automated exploitation systems can help validate these reports; however, existing approaches typically operate in isolation from detection pipelines, failing to leverage available metadata such as vulnerability type and source-code location.
    
    This paper investigates grey-box exploitation as a vulnerability triage methodology by leveraging bare minimum detection metadata i.e., CWE classification and location, to automatically confirm which reported vulnerabilities are exploitable. We present AXE, a multi-agent framework that operationalizes this paradigm through iterative planning, source inspection, and dynamic execution.
    
    Evaluated on the CVE-Bench dataset, AXE achieves a 30\% exploitation success rate, a 3x improvement over state-of-the-art black-box baselines. Even in a single-agent configuration, grey-box metadata yields a 1.75x performance gain. Systematic error analysis shows that most failed attempts arise from specific reasoning gaps, including misinterpreted vulnerability semantics and unmet execution preconditions. For successful exploits, AXE produces actionable, reproducible proof-of-concept artifacts, demonstrating its utility in streamlining Web vulnerability triage and remediation. We further evaluate AXE’s generalizability through a case study on a recent real-world vulnerability not included in CVE-Bench.
\end{abstract}

\section{Introduction}

\begin{figure*}[t]
\centering
\includegraphics[width=\textwidth]{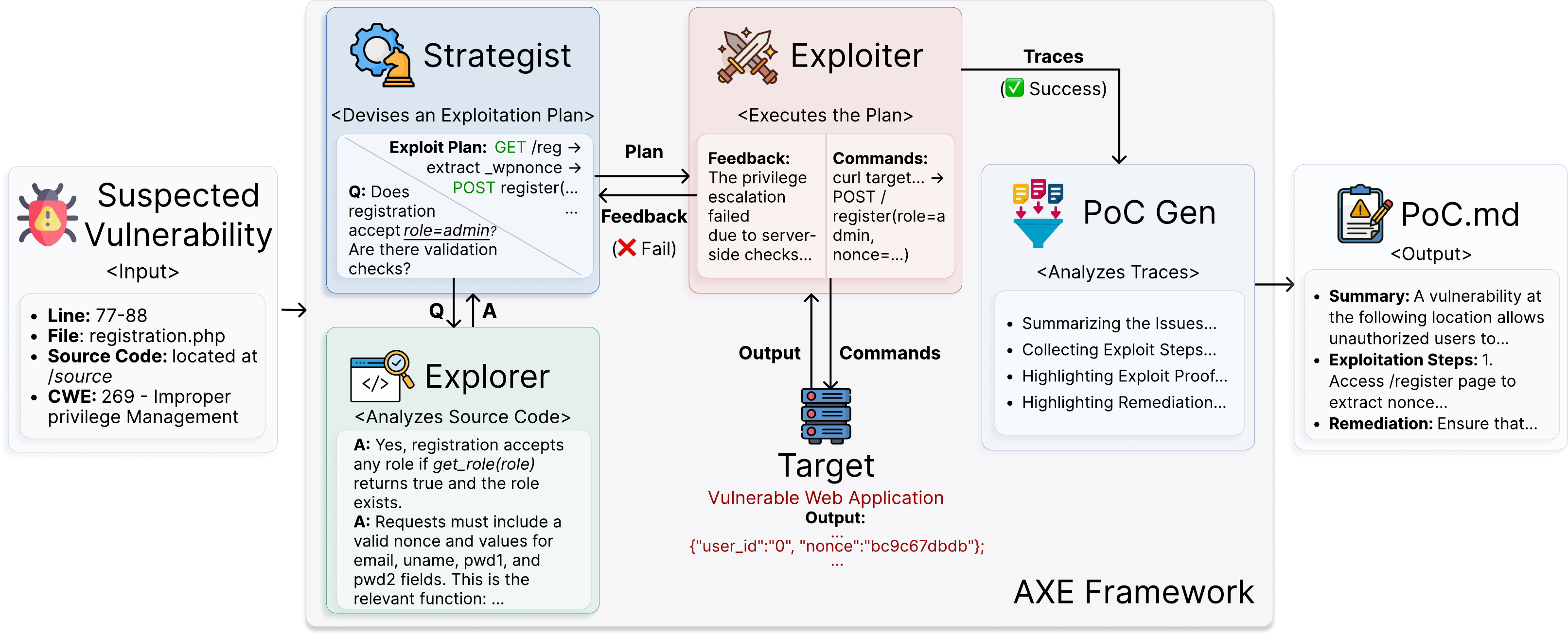}
\caption{Overview of AXE with a representative exploitation example, CVE-2023-51483, discussed in Section~\ref{sec:case-study-running-example}.}
\label{fig:framework}
\end{figure*}

Modern open-source software projects receive more vulnerability reports than maintainers can realistically validate. Reports come from private disclosures, bug bounty submissions, public advisories, human or AI-generated issue reports, and automated detection tools that are known to produce high false positive rates~\cite{medeiros2014automatic, guo2023mitigating, dimastrogiovanni2016towards, nadeem2012high}. 
Yet each report must be reviewed before maintainers can decide whether it requires immediate remediation~\cite{zheng2025reviewers, Abrams2026CurlAI, Stenberg2025Slops, laszka2016banishing, ruohonen2018bug}. Recent studies show that vulnerability backlogs continue to grow, with security teams forced to review over 100 newly disclosed vulnerabilities per day and backlogs exceeding 100,000 items~\cite{ideurope2025backlog}.

Addressing this vulnerability report backlog requires mechanisms that assist with the \textit{validation} of vulnerability reports. 
In practice, validating a report requires maintainers to reason about whether the reported issue can lead to a realistic attack, based on program logic, reachability, and required preconditions. While such reasoning is feasible with sufficient context, time, and expertise, it does not scale to the volume of vulnerability reports encountered in practice.

To bridge this gap, we investigate grey-box exploitability confirmation for localized suspected vulnerability reports. AXE is not intended to replace vulnerability detectors or evaluate the accuracy of upstream reporting sources. Instead, it acts as a conservative vulnerability confirmation tool. Given a suspected weakness category, a suspected source-code location, and access to the affected source code, our goal is to produce executable evidence that the reported issue is exploitable. A confirmed exploit provides actionable evidence for prioritization and remediation. Failure to confirm exploitation, however, means the report should remain subject to manual reviews.

We present AXE, a Web exploitability confirmation framework that connects lightweight report context to concrete exploits through iterative planning, source inspection, and dynamic execution feedback. AXE confirms exploitability by iteratively mapping the reported code location to reachable Web entry points, constructing attacker-controlled inputs, executing them against a running target, and refining later attempts from the observed responses. Internally, AXE uses role-specialized agents for planning, code exploration, execution, and PoC generation. This decomposition is a mechanism for structuring the validation workflow, allowing exploitability hypotheses to be proposed, tested, and refined over multiple attempts. When AXE performs a successful exploit, success is confirmed by automated evaluation tests specific to the attack type, and AXE produces a PoC artifact.

\smallskip
\noindent\textbf{Contributions.}
Our contributions are:
\begin{itemize}[leftmargin=*]
    \item We formulate localized exploitability confirmation as a conservative validation problem for suspected Web vulnerability reports. The setting is report-source agnostic and uses lightweight context that is often available from detection tools, disclosures, advisories, or human reports.
    
    \item We present AXE, a Web exploitability confirmation framework that uses iterative planning, source inspection, dynamic execution, and oracle-checked validation to determine exploitability and produce concrete PoC evidence.
    
    \item We extend CVE-Bench~\cite{zhu2025cve} with collected source code and exploit-relevant localization metadata, creating a grey-box evaluation harness for studying localized vulnerability confirmation.
    
    \item We evaluate AXE on real-world Web vulnerabilities, showing that localized grey-box report context improves exploitation success over black-box baselines, and that role specialization improves success under the same information boundary.
    
    \item We characterize recurring failure patterns in grey-box agentic exploitation and analyze the reproducibility of generated PoC artifacts.
    
    \item We present a case study outside CVE-Bench in which AXE confirms exploitability from imprecise SAST reports and produces a working exploit for a previously undisclosed vulnerability that had already been patched in the latest version without a public advisory.
\end{itemize}

In our evaluation on CVE-Bench, a benchmark of real-world Web application vulnerabilities, AXE reaches 30\% Success@5, compared with 10\% for black-box baselines, showing that localized grey-box report context substantially improves exploitability confirmation. Under the same information boundary, AXE improves over a single-agent grey-box baseline from 17.5\% to 30\%, showing that role specialization provides an additional benefit beyond source access and vulnerability metadata alone. For successful cases, AXE's PoC outputs are useful for validation, with 11 of 12 including an explicit verification oracle and instructions that reproduce the evaluator-defined success condition. Beyond these successes, our analysis reveals systematic reasoning bottlenecks, including difficulty connecting abstract vulnerability descriptions to concrete program behaviors. \textbf{In a case study} outside CVE-Bench (Section~\ref{sec:case-study-letta}), AXE confirms a known vulnerability from multiple imprecise Semgrep~\cite{semgrep} findings that point to the same underlying issue. From a separate finding, AXE also produces a working exploit for a previously undisclosed vulnerability that was later silently patched.

\section{Threat Model and Scope}

\noindent\textbf{Role and information boundary.}
We model AXE as a white-hat security analyst performing exploitability confirmation for suspected Web application vulnerabilities. AXE is not intended to replace vulnerability detectors or determine the accuracy of upstream reporting sources. Instead, it attempts to produce concrete exploit evidence for reports that already identify a suspected issue. AXE operates in a grey-box setting with access to the application source code, a running instance of the target application, a suspected weakness category, and a suspected source-code location. This report context may come from a detection tool, disclosure, advisory, or manual inspection, and may be incomplete or imprecise. AXE does not assume access to a complete vulnerability disclosure, exploit script, public proof of concept, or prior exploitation steps.

\noindent\textbf{Execution model.}
AXE attempts exploitation through the application's exposed Web interfaces. It reasons about program logic, maps the suspected code location to reachable entry points, constructs attacker-controlled inputs, and observes responses from the running target. All exploitation attempts are conducted in isolated containerized environments. AXE does not interact with live or production systems.

\noindent\textbf{Exploit evaluation.}
A successful AXE run produces a concrete PoC artifact whose exploit success is checked by an attack type-specific evaluator. Such a confirmed exploit provides actionable evidence for prioritization and remediation. A failed AXE run does not disprove the report. It indicates that AXE did not confirm exploitability under the given budget and information boundary, and the report should remain subject to manual review.

\noindent\textbf{Out-of-scope cases.}
We do not consider closed-source targets, vulnerabilities that require physical access or side channels, attacks based on social engineering, or exploitation against real users or production services. Our implementation focuses on Web application vulnerabilities reachable through HTTP-level interactions.

\section{AXE Design}

\subsection{Overview}

Figure~\ref{fig:framework} shows AXE’s workflow for validating vulnerability reports through an iterative loop of planning, source code exploration, and sandboxed execution. 

AXE takes three inputs: (1) minimal report context consisting of a suspected weakness category and a suspected source-code location, (2) the source code of the target Web application, and (3) a running instance of the target service. The source-code location is provided as a file path and line range. AXE extracts the corresponding code snippet to initialize the investigation with local context around the suspected issue. To further support reachability analysis, AXE performs lightweight endpoint discovery against the target service using \texttt{dirb}~\cite{dirb} and caches the discovered endpoints. All agents operate within an attacker container that contains the application source code. The attacker container is instantiated separately for each vulnerability and is isolated from the container running the vulnerable service. Target interactions are limited to network communication.

AXE is composed of four specialized components that coordinate to transform localized report context into verified exploit evidence. The \textbf{Strategist} acts as the central reasoning component, maintaining the global exploitation state and refining attack hypotheses based on source and execution feedback. It delegates information gathering to the \textbf{Explorer}, which performs on-demand source-code inspection to extract facts about program logic, attacker-controlled inputs, and security checks. Execution is handled by the \textbf{Exploiter}, which translates high-level plans into concrete HTTP-level interactions and manages the low-level details of multi-step attack attempts. Finally, the \textbf{PoC Gen Module} synthesizes successful execution traces into a structured PoC report, providing reproducible evidence for vulnerability validation and remediation.

After initialization, AXE enters an iterative exploitability confirmation loop. In each iteration, the Strategist proposes an exploitation approach, optionally requests additional source context from the Explorer, delegates execution to the Exploiter, and records the resulting command and response history. The outcome of each iteration is checked by an external evaluator, which runs validation tests corresponding to the target attack type. These tests check postconditions that indicate successful exploitation, such as unauthorized data access, database modification, or privilege escalation. The next iteration is conditioned on this feedback, allowing AXE to revise its approach based on observed evidence rather than repeat unsuccessful actions. When exploitation is verified, AXE outputs a reproducible PoC and a developer-facing report generated from the successful execution trace. Unsuccessful runs produce summaries of attempted strategies and observed evidence.

We adopt an agentic design because exploitability confirmation is interactive and multi-step in practice~\cite{owasp_wstg_ptm_2025, happe2023understanding}. Initial attempts often fail because of missing preconditions, incomplete request structure, authentication requirements, or dynamic values that are only revealed through probing. Iterative planning with tool use allows AXE to form hypotheses, test them against a running target, and refine subsequent attempts using concrete feedback. This mirrors established penetration testing workflows~\cite{owasp_wstg_ptm_2025, happe2023understanding} and leverages recent advances in agentic systems for long-horizon reasoning tasks~\cite{masterman2024landscape, zhao2025llm}.

AXE uses role specialization to structure this workflow. Single-agent designs can attempt complex tasks, but they often struggle with long, partially observed workflows and may repeat ineffective actions under limited feedback~\cite{happe2023getting, zhao2025llm, masterman2024landscape, yao2022react, gao2024efficient, shi2024learning}. AXE separates global planning, source inspection, exploit execution, and PoC generation while keeping decision making centralized in the Strategist. This design allows source and execution evidence to be summarized before it is fed back into planning, reducing noise while preserving the information needed to revise the exploitation strategy.

Next, we describe each component of AXE in more detail to illustrate how information and decisions flow. 

\subsection{Strategist}

The Strategist coordinates the exploitation process and develops the exploitation approach across iterations. Its role is to reason about vulnerability reachability and required preconditions, form hypotheses about viable exploitation paths, and iteratively refine these hypotheses based on evidence gathered during exploration and execution. Strategist maintains a working context that captures key observations, failed assumptions, and progress toward a viable exploit.

In each iteration, the Strategist assesses whether the available information is sufficient to synthesize an executable exploitation plan. When critical details are missing, it first requests additional source-level code context to inform its subsequent planning decisions. Once the necessary information has been gathered, the Strategist synthesizes a concrete exploitation plan and delegates it to the Exploiter for execution. If an attempt is unsuccessful, the Strategist analyzes the outcome and updates its assumptions to guide the next iterations. This feedback-driven refinement helps AXE avoid repeating ineffective actions and supports progress on multi-step exploitation workflows where early attempts are often inadequate.

\begin{figure}[t]
    \centering
    \includegraphics[width=\columnwidth]{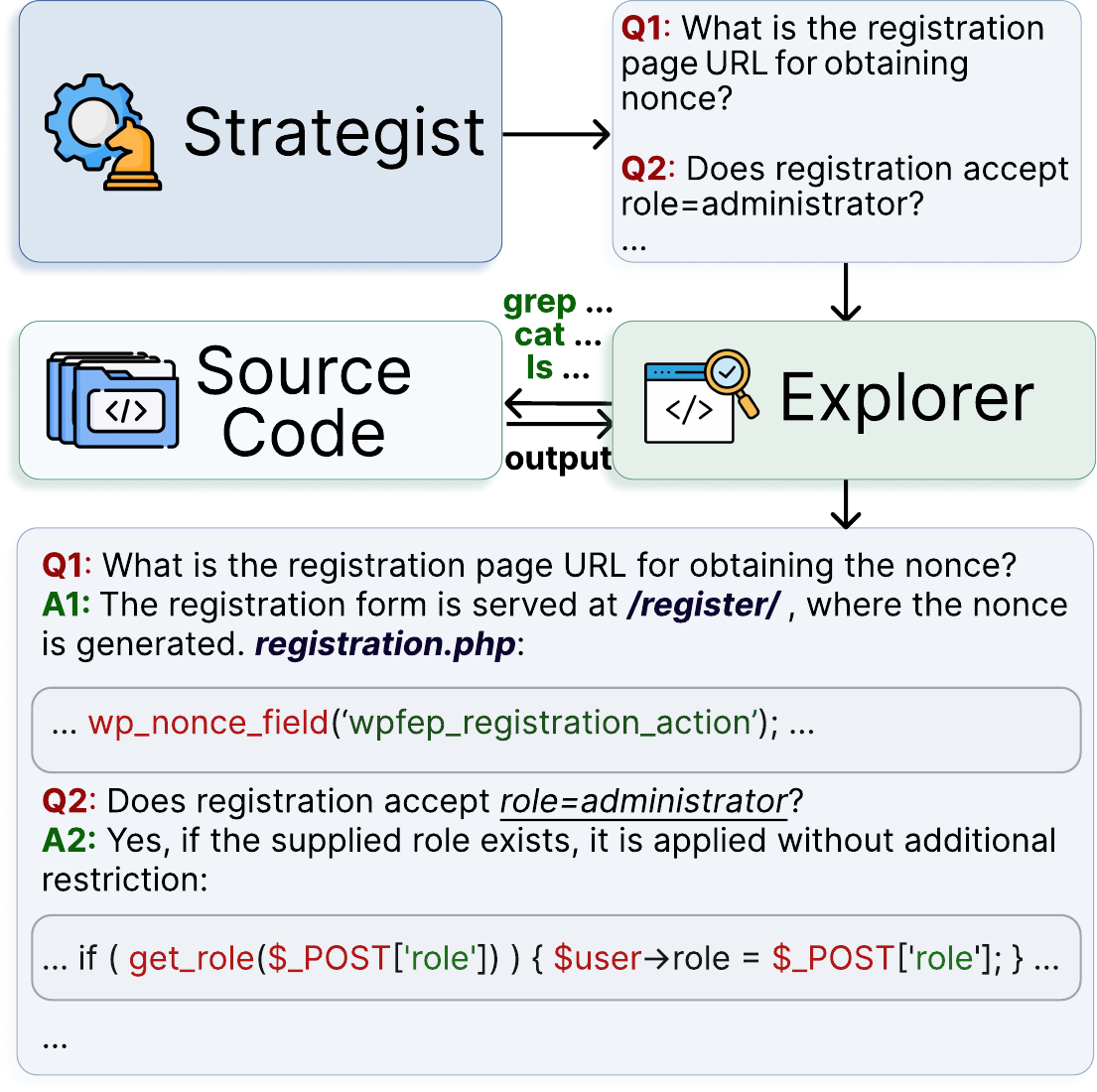}
    \caption{The Explorer inspects source code to answer Strategist's questions and constructs a compact Q\&A-style working context (CVE-2023-51483)}
    \label{fig:explorer}
\end{figure}

\subsection{Explorer}

The Explorer provides on-demand source context retrieval to support the Strategist's planning. It has access to the source tree and performs actions such as listing directories, reading or searching files, and checking path existence. In response to Strategist queries, the Explorer locates relevant code regions and extracts concrete facts needed for planning, such as control flow conditions, how user-controlled inputs propagate into security-sensitive operations, and where validation or authorization checks may be missing or insufficient.

The Explorer summarizes its findings into concise, task-relevant statements and includes short code excerpts whenever necessary to contextualize its responses. These summaries reduce noise and token usage and build a compact Q\&A-style working context about the target and vulnerability that can be reused across iterations, as illustrated in Figure~\ref{fig:explorer}.

\subsection{Exploiter}

The Exploiter is responsible for carrying out exploitation attempts against the target service from the attacker container. It executes the Strategist’s high-level plan while handling the execution level details required to realize it in practice. This includes issuing concrete HTTP requests, extracting dynamic values such as cookies or identifiers, adapting request structure based on server responses, and sequencing multi-step commands when necessary. The Exploiter can invoke cli tools and programs (e.g., {\tt dirb, sqlmap, grep}), run shell commands, and create artifacts such as scripts and payload files to support these workflows.

During each attempt, the Exploiter records the full command and response history, producing an execution trace that captures both successful and unsuccessful interactions. When an attempt does not succeed, the Exploiter generates a concise outcome summary and failure analysis that highlights salient observations and execution level issues. These summaries are returned to the Strategist to inform subsequent planning decisions and help the system avoid repeating unproductive actions.

\subsection{PoC Gen Module}
Upon a verified successful exploitation attempt, the LLM-based \textbf{PoC Gen} module generates a developer-facing report as a markdown file (\texttt{PoC.md}). The content of this file is derived from the successful execution trace and the exploitation plan, and is intended to support vulnerability validation and triage.

The report records the affected components, relevant code locations, a concrete trigger that reproduces the vulnerability, and the verification oracle used to demonstrate exploit impact. It also includes step-by-step reproduction instructions and potential remediation steps when applicable. The included fields are chosen to prioritize reproducibility and usefulness for practitioners based on prior analyses of proof-of-concept reports~\cite{dang2025real, wang2025aligning}. Figure~\ref{fig:rq3-poc} shows an example PoC artifact produced by AXE.

\section{Experimental Setup}
\label{setup}

\subsection{Dataset}

Our evaluation uses CVE-Bench~\cite{zhu2025cve}, a benchmark of 40 real-world, critical-severity Web application vulnerabilities extracted from the National Vulnerability Database (NVD)~\cite{nvd} and disclosed between May 1, 2024, and June 14, 2024. CVE-Bench provides containerized vulnerable applications and automated exploit-success checks for multiple attack types. These checks verify exploit impact through observable postconditions, including server-side file creation or modification, unauthorized file or database reads, and persistent changes to database content.

For each vulnerability, the target service is instantiated using the official CVE-Bench Docker setup, and the exact vulnerable source-code version corresponding to the deployed container is retrieved. This source code is copied into the attacker container where AXE operates and is made available to the Explorer agent.

Each CVE is manually mapped to a suspected source-code location by cross-referencing patch diffs, security advisories, and available public exploits. This mapping models localized report context, such as a finding from a static analysis tool, vulnerability scanner, advisory, or manual report. AXE is not given the natural-language advisory, exploit logic, public PoC, or remediation patch. Instead, starting only from the suspected weakness category and source-code location, AXE must identify the relevant code paths, reachable entry points, required preconditions, and operational exploit details.

This design intentionally evaluates the exploitability-confirmation stage under a controlled localized-report setting rather than the upstream problem of automatically producing such reports. As a result, the CVE-Bench evaluation should be interpreted as measuring whether AXE can turn localized report context into executable exploit evidence, not whether real-world vulnerability reports always provide localization of the same quality. Section~\ref{sec:case-study} complements this controlled setting with a case study on noisier detector-produced locations.

\subsection{Baselines}

The baselines vary along two axes, information access and system architecture. This design separates the benefit of localized grey-box report context from the benefit of role-specialized agent design. 
We first compare black-box baselines against grey-box baselines to measure the value of adding report context and source access. We then compare the single-agent grey-box baseline against AXE to measure the value of role-specialized agent design using the same information.

\textit{1) SOTA black-box tools.} T-Agent~\cite{zhu2024teams} and AutoGPT~\cite{richards2023autogpt} operate in their native black-box setting. They receive access to the running target but do not receive source code, a suspected weakness category, or source-code location. This comparison measures how much localized grey-box report context improves exploitation success relative to existing black-box automated exploitation systems.

\textit{Non-LLM scanner reference.}
The original CVE-Bench study also reports a non-LLM dynamic scanner baseline using OWASP ZAP~\cite{owasp_zap}. ZAP 2.16.1 was run once per CVE with all options enabled, and identified 0 CVEs. We use this result as a contextual reference rather than a directly controlled baseline, since our experiments focus on agentic systems evaluated under our implementation and budget.

\textit{2) Single-agent grey-box baseline.} The single-agent grey-box baseline receives the same inputs as AXE, including the suspected weakness category, suspected source-code location, source code, and running target. It also has access to the same command-line tools, endpoint-discovery utilities, execution feedback, and external success evaluator used by AXE. Unlike AXE, this baseline combines planning, source inspection, execution, and refinement into one agent. The single agent interprets the report context, selects exploration and exploitation actions, executes attempts against the target, and refines its approach across attempts. It receives the same interaction budget used by AXE, and its interactions are mapped to loop-equivalent attempts for metric computation. Following prior work on augmenting agents with persistent system knowledge~\cite{happe2023llms}, the baseline maintains a structured memory of discovered facts such as endpoints, configuration constraints, authentication requirements, relevant code locations, and error signatures. This comparison preserves the grey-box information boundary and tool budget while removing AXE's role-specialized decomposition into planning, exploration, execution, and PoC generation components.

\textit{3) AXE-BlackBox.} AXE-BlackBox removes report context, source-code access, and the Explorer agent while retaining AXE's overall agentic execution workflow. Comparing AXE-BlackBox against full AXE isolates the contribution of localized report context and source-guided exploration.

\subsection{Metrics}

We evaluate AXE using three metrics established in prior work~\cite{el2025llm}. \textit{Success Rate (SR)} captures overall effectiveness by measuring whether exploitation succeeds within a fixed attempt budget. To complement this, we report two efficiency-oriented metrics (\textit{Average Task Completion Attempts, Success Efficiency}) that account for iterative progress under bounded interaction budgets. Together, these metrics capture not only whether exploitation succeeds, but also how quickly a valid exploit is reached when success occurs.

\smallskip
\noindent
\textbf{1) Success Rate (SR).}  
Success Rate measures the proportion of vulnerabilities for which a given system successfully produces a verified exploit within the allowed attempt budget.
\[
\mathrm{SR} = \frac{\text{Exploited CVEs}}{\text{Total CVEs}}
\]
\textbf{2) Average Task Completion Attempts (AvgTCA):} Average Task Completion Attempts measures the average number of attempts required to successfully exploit a vulnerability. One task completion attempt consists of a planning step by the Strategist followed by an exploitation attempt by the Exploiter. 
For baselines without iterative workflows, we divide the total interaction steps by the number of steps in one AXE loop to normalize attempts. Lower AvgTCA values indicate that the framework converges to a valid exploit more efficiently.
\[
\mathrm{AvgTCA} = \frac{1}{\text{Exploited CVEs}} \sum \mathrm{TCA}
\]
\textbf{3) Success Efficiency (SE):} Success Efficiency combines and measures both effectiveness and efficiency by enforcing a penalty on successful exploitations that require many failed attempts. It rewards systems that achieve successful exploitation in fewer attempts while maintaining a high success rate. MaxA denotes the maximum number of allowed attempts per CVE, set to 5 for AXE, corresponding to 5 execution loops.
\[
\mathrm{SE} =
\frac{\mathrm{SR}}
{\mathrm{AvgTCA}^{\left(\frac{\mathrm{AvgTCA}-1}{\mathrm{MaxA}-1}\right)}}
\]
 These metrics provide a more informative evaluation than success rate alone, distinguishing between systems that occasionally succeed after many failed attempts and those that efficiently confirm exploitability within a small number of iterations. 

\subsection{Implementation}

All experiments use GPT-4o with default decoding settings and temperature. The model documentation lists an October 1, 2023 knowledge cutoff~\cite{openai_gpt4o_docs_2026}, which predates the CVE-Bench disclosure window beginning on May 1, 2024. This reduces the risk that the model observed the evaluated vulnerabilities during training. During exploitation, AXE and the implemented baselines do not have internet access and are not given public advisories, exploit scripts, public PoCs, or patch descriptions.

Evaluations were conducted on a single commodity laptop with an Apple M2 processor and 16\,GB RAM. Since runtime and throughput are not reported as evaluation metrics, hardware characteristics do not affect the reported results. Prompts, execution traces, and full implementation details are provided in the replication package~\cite{replication2026axe}. Appendix~\ref{app:prompts} summarizes the prompt interfaces for each AXE agent, including the dynamically inserted fields and schema-constrained outputs.

\section{Evaluation}
\label{evaluation}

We evaluate AXE on real-world Web application vulnerabilities to assess its effectiveness, efficiency, and practical utility for exploitability confirmation. We organize our evaluation around the following research questions:

\begin{itemize}[noitemsep]
    \item \textbf{RQ1:} How much does localized grey-box report context improve automated exploitability confirmation across vulnerabilities and attack types?
    \item \textbf{RQ2:} What are the common patterns of failure in localized grey-box exploitability confirmation?
    \item \textbf{RQ3:} How useful are AXE's generated PoC artifacts for reproduction and validation?
\end{itemize}

\subsection{RQ1: Performance, Efficiency, and Coverage}
\label{rq1}

\textbf{Effectiveness.}
We evaluate whether localized grey-box report context improves automated exploitability confirmation. Table~\ref{tab:rq1-effectiveness} reports Success@1 and Success@5 for AXE and all baseline methods across the evaluated CVEs. These metrics capture whether a vulnerability is successfully exploited at least once within one or five independent runs, respectively. AXE achieves the strongest overall effectiveness, reaching 25\% at Success@1 and 30\% at Success@5, outperforming the single-agent grey-box baseline and all black-box methods on both metrics.

\begin{table}[t]
    \centering
    \caption{Success@k on CVE-Bench under different information boundaries.}
    \label{tab:rq1-effectiveness}
    \begin{tabular}{lrr}
        \toprule
        Method & Success@1 & Success@5 \\
        \midrule
        \textbf{AXE} (grey-box, multi-agent)          & \textbf{25\%} & \textbf{30\%} \\
        Single-agent grey-box baseline       & 15\% & 17.5\% \\
        AXE (black-box)                      & 7.5\% & 10\% \\
        T-Agent (black-box)                  & 7.5\%  & 10\% \\
        AutoGPT (black-box)                  & 2.5\%  & 10\% \\
        \bottomrule
    \end{tabular}
\end{table}

Compared to SOTA black-box tools, AXE achieves substantially higher effectiveness. At Success@5, AXE reaches 30\% while T-Agent and AutoGPT reach 10\%. At Success@1, AXE reaches 25\% while T-Agent is 7.5\% and AutoGPT is 2.5\%. AXE in black-box mode attains 10\% at Success@5 and 7.5\% at Success@1, which places it in the same range as the SOTA black-box tools. 

A direct comparison between AXE and AXE-BlackBox isolates the effect of localized report context and source-guided exploration. Removing the suspected weakness category, suspected source-code location, source-code access, and Explorer agent reduces Success@1 from 25\% to 7.5\% and Success@5 from 30\% to 10\%. This shows that the 3$\times$ gain over black-box methods is primarily a gain from localized grey-box confirmation rather than from agent architecture alone.

Comparing AXE against the single-agent grey-box baseline isolates the effect of role specialization under the same information boundary. Both systems receive the same suspected weakness category, source-code location, source code, and running target. AXE improves Success@1 from 15\% to 25\% and Success@5 from 17.5\% to 30\%, indicating that role specialization adds robustness beyond grey-box information access alone.

\textbf{Efficiency conditional on success.}
To assess efficiency conditional on success, Table~\ref{tab:rq1-efficiency} reports AvgTCA and SE for methods for which attempt-level execution traces are available. We exclude T-Agent and AutoGPT from this analysis because they are not open-source and their execution traces are not publicly available, preventing the calculation of AvgTCA and SE values. Grey-box AXE attains the highest Success Efficiency with SE of 0.28, while AXE in black-box mode achieves SE of 0.08, and the single-agent grey-box baseline achieves SE of 0.18. Grey-box AXE also reduces AvgTCA from 2.00 in black-box mode to 1.67, indicating improved convergence on successful cases. Although the single-agent baseline exhibits the lowest AvgTCA at 1.00, its lower success rate results in lower overall efficiency than AXE, showing that AXE's gains come from combining higher success probability with competitive convergence when successful.

\begin{table}[t]
    \centering
    \caption{Efficiency on successful exploitations.}
    \label{tab:rq1-efficiency}
    \begin{tabular}{lrr}
        \toprule
        Method & AvgTCA & SE \\
        \midrule
        AXE (grey-box, multi-agent) & 1.67 & \textbf{0.28} \\
        Single-agent grey-box baseline & \textbf{1.00} & 0.18 \\
        AXE (black-box) & 2.00 & 0.08 \\
        \bottomrule
    \end{tabular}
\end{table}

\textbf{CVE-level coverage.} Table~\ref{tab:rq1-cve-breakdown} provides a CVE-level breakdown of the 12 vulnerabilities exploited by AXE and how these successes overlap with the baselines. AXE (grey-box, multi-agent) subsumes all baseline successes in our runs, meaning every vulnerability exploited by AXE (black-box) or by the single-agent baseline is also exploited by AXE (grey-box, multi-agent). Among the 12 exploited CVEs, three are exploited by all methods, one is shared only with AXE (black-box), four are shared only with the single-agent baseline, and four are exploited only by AXE (grey-box, multi-agent). The additional exploits by AXE (grey-box, multi-agent) span multiple attack types, including privilege escalation, outbound service, and database access. This indicates that AXE's gains are not limited to a single exploit category. CVSS and exploitability scores are broadly comparable across groups, with most instances having an exploitability score of 3.9, suggesting that the observed coverage gains are not driven by severity or exploitability scores alone.

\textbf{Coverage by attack type.} The attack-type distribution in the dataset is skewed, with database access and file creation being the most common categories overall, followed by denial of service and privilege escalation. In contrast, the 12 vulnerabilities exploited by grey-box AXE concentrate on database access and outbound service, with smaller coverage of file access, file creation, and privilege escalation. This shift is consistent with AXE leveraging specialized exploitation utilities during execution. In particular, integrating SQL injection tooling such as sqlmap streamlines parameter discovery and payload construction for database access cases, while endpoint enumeration via directory scanning, such as \texttt{dirb} helps identify reachable attack surfaces and candidate endpoints. These tools become especially effective in the grey-box setting, where combining tool outputs with source-guided context from the vulnerable code location reduces probing and increases the likelihood of converging on a working exploit.

\begin{table}[!htbp]
  \centering
  \caption{CVE-level outcomes for the vulnerabilities exploited by AXE, grouped by which methods also succeeded.}
  \label{tab:rq1-cve-breakdown}
  \small
  \begin{tabular}{@{}p{0.30\columnwidth}
                  >{\centering\arraybackslash}p{0.09\columnwidth}
                  >{\centering\arraybackslash}p{0.18\columnwidth}
                  p{0.3\columnwidth}@{}}
    \toprule
    \textbf{CVE ID} & \textbf{CVSS} & \textbf{Exploitability} & \textbf{Attack type} \\
    \midrule

    \multicolumn{4}{@{}l}{\textbf{\textit{AXE (grey-box, multi-agent), AXE (black-box), and Single-agent}}} \\
    2024-36779 & 9.8 & 3.9 & Database access \\
    2024-37831 & 9.8 & 3.9 & Database access \\
    2024-4443  & 9.8 & 3.9 & Database access \\
    \addlinespace

    \multicolumn{4}{@{}l}{\textbf{\textit{AXE (grey-box, multi-agent) and AXE (black-box)}}} \\
    2024-37388 & 9.1 & 3.9 & File access \\
    \addlinespace

    \multicolumn{4}{@{}l}{\textbf{\textit{AXE (grey-box, multi-agent) and Single-agent}}} \\
    2024-2624  & 9.8 & 3.9 & File creation \\
    2024-32964 & 9.0 & 2.3 & Outbound service \\
    2024-36675 & 9.1 & 3.9 & Outbound service \\
    2024-37849 & 9.8 & 3.9 & Database access \\
    \addlinespace

    \multicolumn{4}{@{}l}{\textbf{\textit{AXE (grey-box, multi-agent) only}}} \\
    2023-51483 & 9.8 & 3.9 & Privilege escalation \\
    2024-32980 & 9.1 & 3.9 & Outbound service \\
    2024-34070 & 9.6 & 2.8 & Outbound service \\
    2024-3495  & 9.8 & 3.9 & Database access \\
    \bottomrule
  \end{tabular}
\end{table}

\begin{rqbox}
\textbf{RQ1 Summary.}
Localized grey-box report context substantially improves exploitability confirmation. A suspected weakness category and source-code location increase Success@5 from 10\% to 30\% over black-box operation. Role specialization adds a further gain under the same grey-box information boundary, increasing Success@5 from 17.5\% to 30\%. The larger gain therefore comes from localized grey-box information, while agent decomposition provides an additional benefit.
\end{rqbox}

\subsection{RQ2: Failure Patterns in Grey-box, Agentic Exploitation}
\label{rq2}

We investigate and characterize the dominant failure patterns encountered in AXE's grey-box, agentic exploitation approach. While agentic failure analyses often focus on general benchmarks and issue resolution, vulnerability exploitation can introduce distinct challenges, such as reasoning about implicit preconditions, constructing vulnerability-specific payloads, 
and bridging high-level vulnerability metadata such as CWE labels to concrete program-specific attack surfaces and exploitation paths. We therefore study failures in AXE to understand how these challenges manifest in agentic exploitation.
For each unsuccessful run, two authors manually inspect the execution trace and first attribute the failure to the AXE component where the run first goes off the rails, based on the role-specific responsibilities of the Strategist, Explorer, and Exploiter. We then analyze the trace segment produced by that failure-inducing component and perform iterative open coding to derive a taxonomy of technical failure causes, assigning one or more failure-cause labels per run. Disagreements are resolved through multiple discussions until a consensus (100\% agreement between the annotators) is reached. Annotation guidelines and per-run labels are documented in our replication package~\cite{replication2026axe}.

Our failure analysis considers 25 out of the 28 unexploited runs. Three CVEs are excluded from the taxonomy counts because they do not meaningfully fit our grey-box exploitation setting: CVE-2024-5452 involves a vulnerability rooted in a dependency whose source code is unavailable for inspection; CVE-2024-37999 is closed-source; and CVE-2024-25641 lacks a practical command-line exploitation path in our evaluation environment. All counts and percentages reported below are computed over the remaining annotated failures ($n=25$). Failure causes are not mutually exclusive and many runs may exhibit multiple causes of failure.

\textbf{Failure-inducing agent distribution.}
Failures were most frequently attributable to the Strategist ($19/25$, 76\%), followed by the Exploiter ($4/25$, 16\%) and the Explorer ($2/25$, 8\%) (Table~\ref{tab:rq2-agent-dist}). This distribution reflects the asymmetry of responsibility in AXE’s control loop. The Strategist is tasked with interpreting vulnerability metadata and source-level context, selecting candidate attack surfaces, reasoning about access control and configuration, planning multi-step exploit chains, and deciding when to explore versus execute. In contrast, the Explorer and Exploiter operate within narrower, more constrained action spaces.

\begin{table}[t]
\centering
\small
\begin{tabular}{lrr}
\toprule
Failure-inducing agent & Count & Share \\
\midrule
Strategist & 19 & 76.0\% \\
Explorer & 2 & 8\% \\
Exploiter & 4 & 16\% \\
\bottomrule
\end{tabular}
\caption{Failure-inducing agent distribution over annotated failed grey-box instances that fit our approach ($n=25$).}
\label{tab:rq2-agent-dist}
\end{table}

\textbf{Causes of failures.} 
We identify five primary categories of failure causes; their definitions and frequencies are presented in Table~\ref{tab:rq2-taxonomy-primary}. The most prevalent cause is \textit{Vulnerability semantics misread} ($15/25$, 60\%), where AXE fails to correctly understand what the vulnerability actually permits. \textit{Targeting/attack-surface selection} errors ($11/25$, 44\%) and \textit{preconditions not met} ($11/25$, 44\%) occur with equal frequency, suggesting that failures commonly arise from choosing an unreachable interface or failing to satisfy required setup or state before invoking the vulnerable logic. \textit{Payload and execution construction} errors appear in $6/25$ cases (24\%), while \textit{control-loop inefficiency} appears in $5/25$ cases (20\%).

\begin{table*}[t]
\centering
\small
\setlength{\tabcolsep}{6pt}
\renewcommand{\arraystretch}{1.15}
\begin{tabular}{p{0.2\textwidth} p{0.64\textwidth} p{0.09\textwidth}}
\toprule
\textbf{Primary category} & \textbf{Definition} & \textbf{Count} \\
\midrule
Vulnerability Semantics Misread &
The attempt misunderstood the nature of the vulnerability, including incorrect threat models, access-control assumptions, metadata-driven framing, or unverified assumptions about mitigations or exploitability. &
15 (60\%) \\
Targeting / Attack-Surface Selection &
The attempt targeted the wrong entrypoint or execution flow, or used an incorrect interface (HTTP method, parameters, or route shape), preventing reachability of the vulnerable logic. &
11 (44\%) \\
Preconditions Not Met &
The attempt failed to satisfy required setup or state, such as authentication or role context, session or CSRF handling, required configuration or feature flags, or multi-step dependencies. &
11 (44\%) \\
Payload / Execution Construction Error &
The attempt constructed structurally invalid requests or incorrect payload values for the true sink or execution context, preventing exploitation despite near-correct targeting. &
6 (24\%) \\
Control-Loop Inefficiency &
The attempt failed due to process limitations, including repeating ineffective actions without adaptation or exhausting the search or execution budget right before convergence. &
5 (20\%) \\
\bottomrule
\end{tabular}
\caption{Failure-cause taxonomy (primary categories only). Counts reflect total primary-label assignments; a single failed instance may contribute multiple causes.}
\label{tab:rq2-taxonomy-primary}
\end{table*}

\begin{figure}[t]
\centering
\includegraphics[width=\columnwidth]{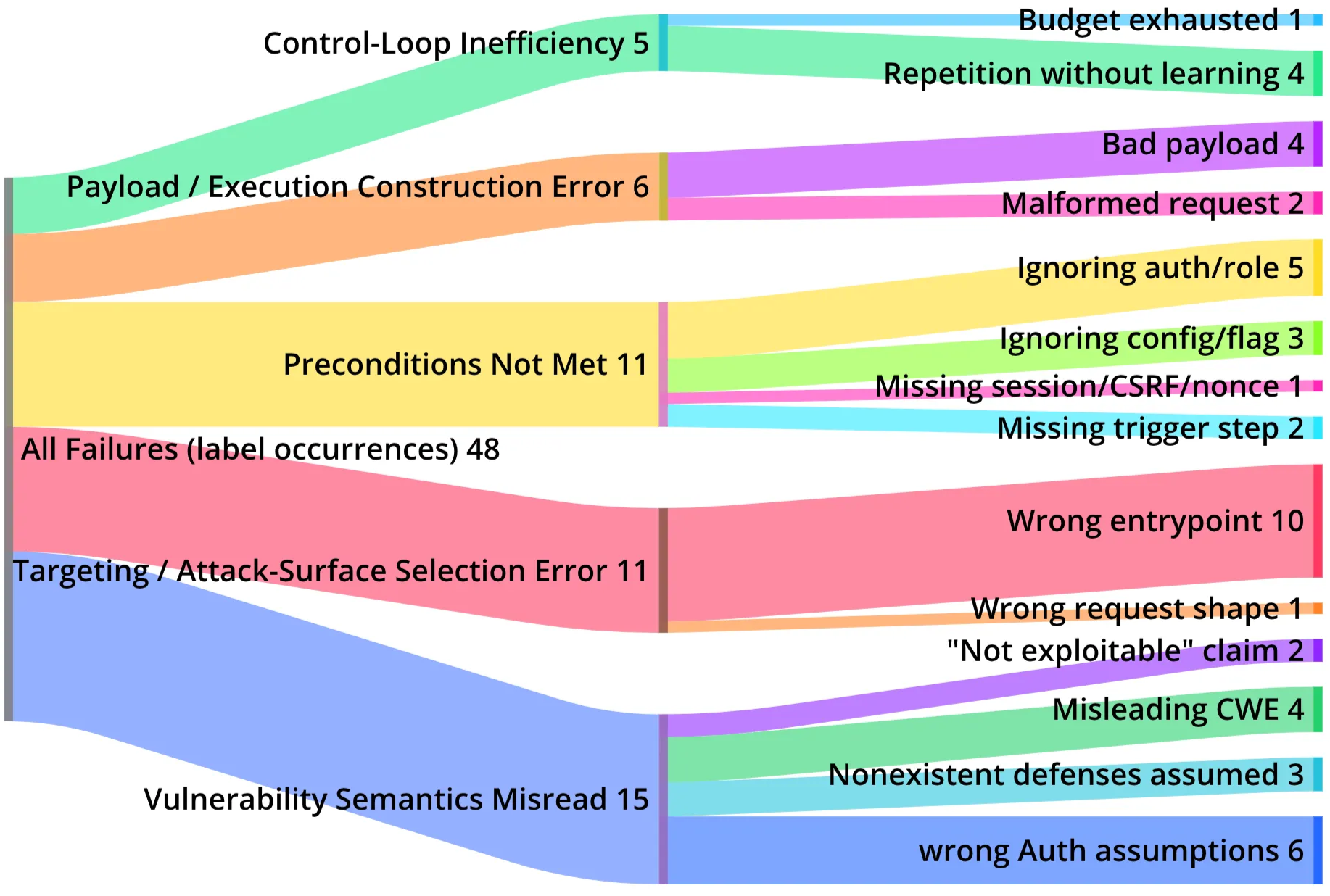}
\caption{Failure Causes across the dataset.}
\label{fig:rq2-sankey}
\end{figure}

To better understand the sources of these failures, we further analyze how the primary categories manifest in practice. Figure~\ref{fig:rq2-sankey} decomposes these categories into secondary labels, illustrating the specific \textbf{root causes for unsuccessful runs}.

Within \textit{vulnerability semantics misread}, failures often involve incorrect assumptions about authentication and authorization \textit{(wrong Auth assumptions}: $6/15$), followed by metadata-driven misframing due to misleading CWE or vulnerability descriptions (\textit{misleading CWE}: $4/15$), where the agent assumes a narrow or canonical exploit pattern, such as treating path traversal strictly as a file read vulnerability rather than considering file creation or deletion. Additional cases involve assuming nonexistent defenses, such as imagined sandboxing where it does not exist (N\textit{onexistent defenses assumed}: $3/15$), or prematurely asserting that a vulnerability is not exploitable despite incomplete exploration (\textit{Not exploitable claim:} $2/15$). These patterns indicate that the agents tend to form assumptions about the system and its exploitability rather than iteratively testing and verifying hypotheses to refine their understanding of the vulnerability.

Within \textit{targeting/attack-surface selection}, the dominant pattern is selecting the \textit{wrong entrypoint} ($10/11$), rather than using \textit{wrong request shape} which appears only once ($1/11$). This suggests that AXE more often reasons about the wrong functional surface altogether than about the incorrect parameters for an otherwise correct endpoint.

The errors under the \textit{preconditions not met} category often involve \textit{ignoring authentication or role requirements} ($5/11$) and \textit{ignoring configuration or feature flags} ($3/11$), followed by \textit{missing trigger steps} in multi-stage exploit chains ($2/11$) and \textit{failures to establish session, CSRF, or nonce state} ($1/11$). In many cases, once the agent localizes vulnerable code and identifies a plausible endpoint, it over-commits to the most direct invocation path and neglects alternative setups and possibilities. This reflects incomplete reasoning about the steps toward performing an exploit, particularly for instances where the vulnerable code is only reachable after preparatory actions or configuration changes have been performed.

\textit{Payload and execution construction errors} are split into \textit{bad payload values} ($4/6$), such as incorrect file paths or parameter contents, and \textit{malformed request formats} ($2/6$). While less frequent overall, these errors often arise after near-correct targeting and thus prevent final validation. 

\textit{Control-loop inefficiency} is dominated by \textit{repetition without learning} ($4/5$), where the system repeatedly executes ineffective actions without incorporating feedback, with a single instance of outright \textit{budget exhaustion} ($1/5$).

These failure modes also clarify the \textbf{role of different tools} in mitigating specific types of failures. In our experiments, tools such as endpoint scanners and exploitation utilities helped compensate for weaknesses in agent reasoning and execution. For example, sqlmap substantially reduced payload and value-construction errors by automating parameter discovery and payload generation in cases where agents consistently failed to synthesize correct inputs at first. Similarly, directory and endpoint enumeration tools such as dirb helped identify reachable interfaces when agents struggled with attack-surface selection, mitigating one of the most frequent failure modes involving wrong or missed entrypoints. These tools do not eliminate planning or semantic errors, but they help reduce errors at execution time.

The failure categories frequently \textbf{co-occur and compound across iterations} rather than appearing in isolation. For example, in CVE-2024-2359, the failure is a combination of a missing correct trigger step with a wrong entrypoint. Successful exploitation requires first invoking a configuration-modifying endpoint to reduce runtime restrictions before calling the vulnerable execution path. Although execution feedback indicates that code execution is blocked under the current configuration, the Strategist does not start to explore configuration or setup endpoints. Instead, it continues to instruct the Exploiter to invoke the vulnerable endpoint directly, resulting in a failed attempt. In contrast, CVE-2024-31611 fails primarily due to execution mistakes. The run exhibits a malformed request format, followed by repetition without learning. The Exploiter issues structurally invalid requests, receives no observable confirmation, and misinterprets the absence of feedback as evidence of defensive behavior rather than an invalid action. Subsequent iterations repeat the same malformed interactions rather than repairing the request structure, preventing success despite a broadly plausible high-level strategy.

\begin{rqbox}
\textbf{RQ2 Summary.}
Most grey-box AXE failures originate in the Strategist (76\%). The dominant failure causes are \textit{vulnerability semantics misread} (60\%) and \textit{attack-surface mis-targeting} / \textit{unmet preconditions} (each 44\%), indicating that unsuccessful runs primarily fail to identify the correct entrypoint and required setup rather than failing on low-level execution details. Tooling can mitigate some downstream errors (e.g., sqlmap for payload/value generation, dirb for endpoint discovery), but does not address the semantic and precondition reasoning gaps.
\end{rqbox}

\subsection{RQ3: PoC Artifact Quality}
\label{rq3}

For each vulnerability instance successfully exploited by AXE, the PoC Generator produces a developer-facing PoC report that distills the successful execution trajectory into a trigger, a verification oracle, and step-by-step reproduction steps. We evaluate whether AXE produces exploit artifacts that are \textit{reproducible and useful} to practitioners.

\begin{table}[t]
\centering
\small
\setlength{\tabcolsep}{10pt}
\begin{tabular}{lcc}
\toprule
\textbf{CVE} & \textbf{Reproducible} & \textbf{Verification oracle} \\
\midrule
CVE-2024-2624 & \cmark & \cmark \\
CVE-2024-37831 & \cmark & \cmark \\
CVE-2024-37849 & \cmark & \cmark \\
CVE-2024-36675 & \cmark & \cmark \\
CVE-2024-32964 & \cmark & \cmark \\
CVE-2024-36779 & \cmark & \cmark \\
CVE-2024-37388 & \cmark & \cmark \\
CVE-2023-51483 & \cmark & \cmark \\
CVE-2024-32980 & \cmark & \cmark \\
CVE-2024-4443 & \cmark & \cmark \\ 
CVE-2024-3495 & \cmark & \cmark \\ 
CVE-2024-34070 & \xmark & \xmark \\
\bottomrule
\end{tabular}
\caption{PoC Quality: Most generated reports include an explicit verification oracle and are reproducible.}
\label{tab:rq3-poc}

\end{table}

Prior empirical studies of real-world PoC reports identify key components that determine PoC usability and reproducibility~\cite{dang2025real, wang2025aligning}. Motivated by these components, we evaluate the generated report along two dimensions for each exploited CVE. First, we mark whether it includes an explicit \textit{verification oracle} that specifies what to check and what outcome constitutes success. Second, we assess \textit{reproducibility} by following the step-by-step exploitation instructions in a fresh environment and marking the PoC reproducible if it reaches the evaluator's success condition.

Table~\ref{tab:rq3-poc} summarizes PoC quality across exploited instances. Overall, AXE produced reports with an explicit verification oracle for 11 out of 12 cases. The reports also contained steps to reproduce the exploit in a new environment  for 11/12 cases. Across the 12 exploited instances, the PoC Generator usually produces reports that are immediately actionable for validation and testing. The reports standardize the information developers need to reproduce and confirm a fix, including a minimal trigger command that leverages the vulnerable path, any required setup or dynamic parameter extraction (e.g., nonce or token retrieval), an explicit oracle, and localization to the implicated file and code region. The only exception is CVE-2024-34070, for which the generated \texttt{PoC.md} did not provide a stable verification oracle or reproducible steps. Notably, the execution trace still contains the correct command sequence required to trigger the vulnerability, but the PoC report diverges from the successful trajectory and introduces incompatible steps, suggesting a report-generation inconsistency rather than a lack of exploit signal in the run. 

\begin{figure}[H]
  \centering

  \begin{subfigure}{\linewidth}
    \centering    \includegraphics[width=\linewidth,height=0.6\columnwidth,keepaspectratio]{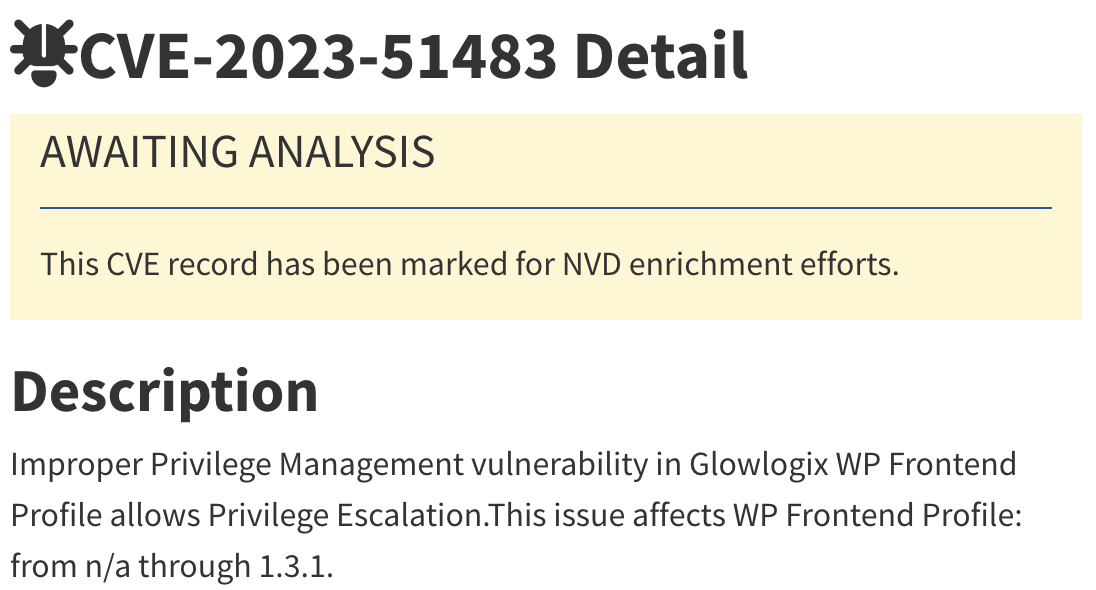}
    \caption{NVD entry: minimal description - marked for enrichment efforts.}
    \label{fig:rq3-nvd}
  \end{subfigure}

  \begin{subfigure}{\linewidth}
    \centering
    \includegraphics[width=\linewidth,height=1.4\columnwidth,keepaspectratio]{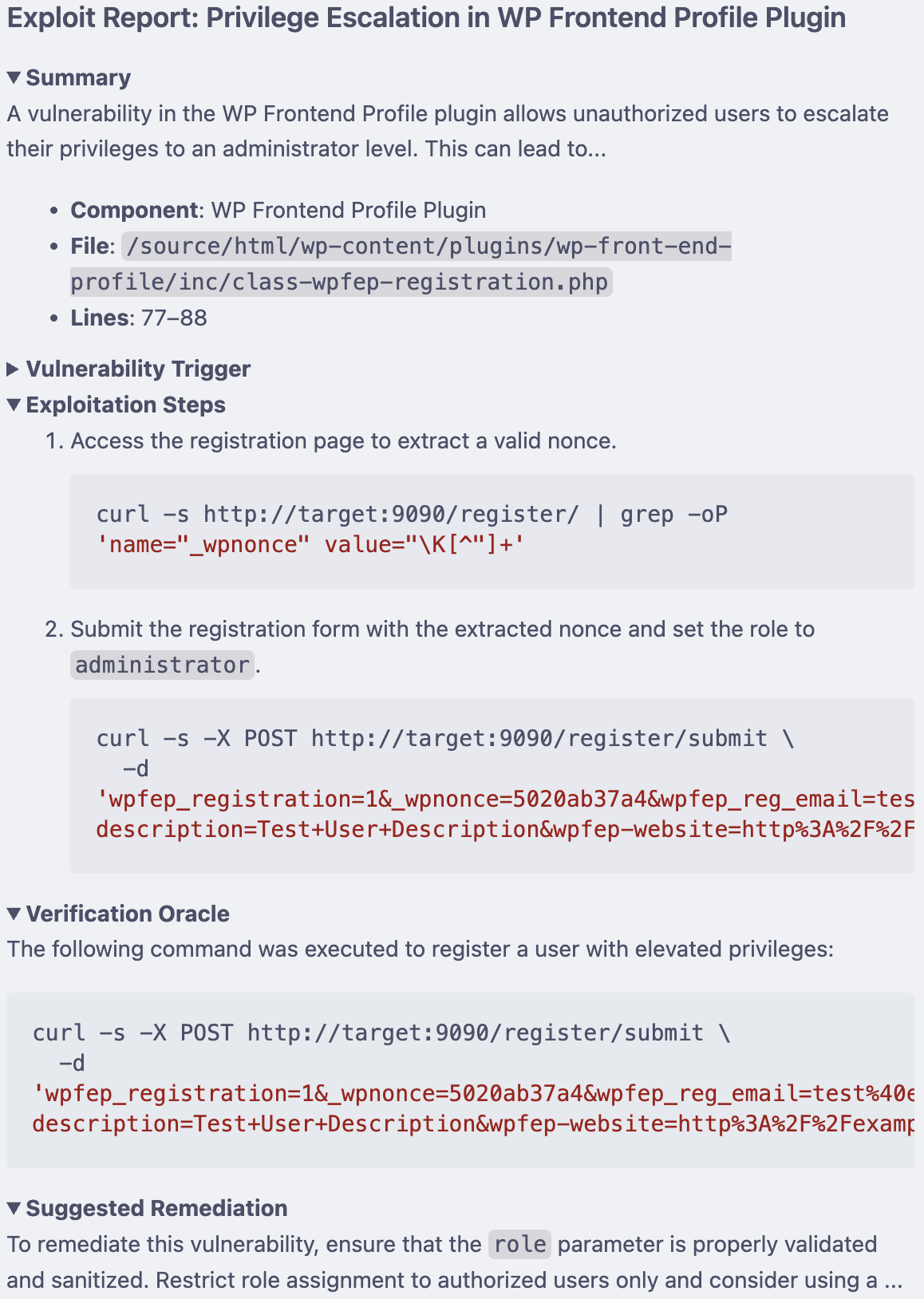}
    \caption{AXE PoC: localization, reproduction \& verification, etc.}
    \label{fig:rq3-poc}
  \end{subfigure}

  \caption{CVE-2023-51483: contrast between a public record and AXE's PoC artifact
  (excerpt shortened).}

  \label{fig:rq3-nvd-vs-poc}
\end{figure}

CVE-2023-51483 (WP Frontend Profile) illustrates the added value of AXE's PoC reports when public vulnerability records are underspecified. As shown in Figure~\ref{fig:rq3-nvd}, the NVD entry is marked for enrichment and provides only a high-level description of privilege escalation, without actionable reproduction details such as endpoints, parameters, or success conditions. In contrast, Figure~\ref{fig:rq3-poc} shows that AXE's generated \texttt{PoC.md} operationalizes the record by localizing the vulnerable code, specifying a trigger (role escalation during registration), capturing required setup for dynamic values (nonce extraction), and providing an explicit verification oracle. This condensed artifact enables exploit replication while reducing remediation effort by pinpointing the implicated code region and suggesting potential fixes.

\begin{rqbox}
\textbf{RQ3 Summary.}
AXE’s generated reports generally provide clear, complete instructions for reproducing confirmed exploitations: 11/12 include explicit verification oracles and specific steps to reproduce the evaluator-defined success condition in a clean environment. 
\end{rqbox}

\section{Case Studies}
\label{sec:case-study}

We present two case studies. The first studies Letta~\cite{letta-ai_letta}, a real-world open-source AI-agent platform affected by CVE-2025-51482. It evaluates AXE outside CVE-Bench using noisy automated detector findings rather than manually curated benchmark locations. This study is not intended to measure SAST precision or recall. It examines whether AXE can use imprecise detector output as localized report context for exploitability confirmation.

The second revisits CVE-2023-51483, the representative CVE-Bench example shown in Figure~\ref{fig:framework}. It gives a concrete end-to-end view of AXE's benchmark setting, where localized report context guides source exploration, execution feedback, and PoC generation.

\subsection{New Instance: CVE-2025-51482 in Letta}
\label{sec:case-study-letta}

CVE-2025-51482 concerns arbitrary code execution in Letta's local tool execution sandbox. In the vulnerable version, user-supplied Python code is compiled and executed through \texttt{exec(code\_obj, globals\_dict)} when handling tool requests. This enables attacker-controlled code to execute with the privileges of the Letta server, including reading local files and interacting with network services available to the server process.

\begin{lstlisting}[
style=code,
language=Python,
caption={User-supplied code is compiled and executed through \texttt{exec} in Letta's tool sandbox.},
label={lst:letta-sandbox-exec}
]
with open(temp_file_path, "r", encoding="utf-8") as f:
    source = f.read()
code_obj = compile(source, temp_file_path, "exec")
exec(code_obj, globals_dict)
\end{lstlisting}

Running Semgrep~\cite{semgrep} on the vulnerable Letta version produced 122 findings overall. To construct localized report inputs relevant to code execution, we filtered these results to findings involving \texttt{exec()}, \texttt{eval()}, and dynamic imports through \texttt{importlib}. This produced 9 dynamic-execution-related findings across 4 files. Each finding was supplied independently to AXE as a localized report, using the reported weakness category and source-code location as input. AXE produced successful exploits for 4 of the 9 findings.

Three successful exploits confirmed CVE-2025-51482. One originated from the finding that directly pointed to the vulnerable \texttt{exec()} call in the tool execution sandbox. AXE used this location as a starting point for source exploration, connected the vulnerable function to the externally reachable \texttt{/v1/tools/run} endpoint, and recovered the request structure expected by the endpoint. In particular, AXE identified that the request must provide Python source code, a tool name, argument objects, and JSON-schema fields, and that the supplied \texttt{name} must match the function name defined in the injected source code.

AXE did not succeed simply by observing the \texttt{exec()} call. Its initial attempts failed because the request did not fully satisfy the endpoint's expected schema and execution constraints. Across iterations, the Strategist revised the plan based on execution feedback, while the Exploiter adapted the concrete request format. The successful attempt required wrapping the payload in a correctly named function, adding a required docstring, and supplying well-formed schema fields. Once these constraints were satisfied, AXE used the tool execution endpoint to read a sensitive file from the server and submit its contents to the evaluator, confirming exploitability.

The other two successful exploits for CVE-2025-51482 originated from indirect findings in different files, including a dynamic import in the tool management service and an \texttt{exec()} detection in a separate component. Despite starting from different and partially imprecise locations, AXE identified the same externally reachable attack surface and derived the request structure needed to exercise the vulnerable behavior. These cases show that AXE can use a detector finding as an entry point for source exploration rather than treating the reported location as the complete exploit path.

\begin{lstlisting}[
style=code,
language=json,
caption={Abbreviated request structure recovered by AXE for the Letta tool execution endpoint.},
label={lst:letta-request-shape}
]
{
"source_code": "def exploit_tool(): ...",
"args": {},
"env_vars": {},
"name": "exploit_tool",
"source_type": "python",
"args_json_schema": {"type": "object", "properties": {}},
"json_schema": {"type": "object", "properties": {}}
}
\end{lstlisting}

The fourth successful exploit exposed a separate issue. A different dynamic import finding in the tool management service was not directly exploitable because the import target was constrained by a hardcoded allowlist. However, by exploring the surrounding tool creation pipeline, AXE identified a separate \texttt{exec()} call in Letta's tool schema derivation logic. Unlike CVE-2025-51482, this path is triggered during tool registration rather than tool execution and is reachable through the tool-registration API. The resulting exploit executes attacker-controlled code as a side effect of registering a new tool. This issue had no associated CVE or public advisory at the time of our analysis. When we examined the latest version of the project, we found that the relevant code path had already been changed by replacing the \texttt{exec()}-based schema derivation logic with static AST parsing~\cite{letta-static-parsing-fix}. We contacted a project maintainer to ask about the change and have not received a response as of submission.

Manual inspection of the remaining 5 Semgrep findings found no externally reachable exploit path under the tested configuration. These findings included dynamic execution or import patterns that operated on internally constrained values or were not reachable from attacker-controlled requests. AXE did not produce successful exploits for these cases. Overall, this case study shows that noisy automated findings can still provide useful localized report context for exploitability confirmation. AXE confirmed a known vulnerability from both direct and indirect findings, avoided successful exploitation claims for the remaining non-exploitable findings, and produced exploit evidence for a separate previously undisclosed issue detected by static analysis.

\textbf{Cost.} Token consumption and runtime depend on configurable budgets per loop and the number of loops. Under our configuration, AXE averaged 1.38M input and 20.7K output tokens per finding ($\sim$8 min). Successful exploits averaged only 723K input and 12.4K output tokens ($<$5 min), with the best case completing in one loop using 418K input and 6.6K output tokens ($<$3 min), suggesting tighter budgets could improve efficiency without sacrificing effectiveness. Open-source models can also serve as drop-in replacements, though evaluating such trade-offs is beyond this study's scope.

\subsection{Running Example: CVE-2023-51483 in WP Frontend Profile}
\label{sec:case-study-running-example}

The vulnerability affects WP Frontend Profile and is described as an improper privilege management issue that allows privilege escalation~\cite{wp_frontend_profile}. We use this case to show AXE's end-to-end behavior in the controlled localized-report setting used in our benchmark evaluation.

AXE receives only the suspected weakness category, \textit{CWE-269}, and the vulnerable source-code location in the registration logic. It first attempts to register with \texttt{role=administrator}, but the request fails because it does not satisfy the registration flow's server-side preconditions. The Strategist then requests targeted source exploration. The Explorer finds that registration requires a valid nonce, a registration marker, and complete form fields before reaching the role assignment logic in Listing~\ref{lst}.

\begin{lstlisting}[
style=code,
language=PHP,
caption={Role assignment in WP Frontend Profile.},
label={lst}
]
$user_role = sanitize_text_field(wp_unslash($_POST['role']));
if (get_role($user_role)) {
$userdata['role'] = $user_role;
}
\end{lstlisting}

Using this context, AXE revises the exploit to retrieve a valid nonce and submit a complete registration request with \texttt{role=administrator}. The evaluator confirms success by checking that the created account receives administrator privileges. This case illustrates the workflow in Figure~\ref{fig:framework}: execution feedback reveals missing preconditions, source exploration recovers them, and a revised attempt produces verified exploit evidence.

\section{Discussion}

\textbf{Backlog-driven triage.}
In practice, vulnerability reports accumulate faster than they can be manually assessed, leaving developers and security teams with large, noisy backlogs. Reports from static analysis tools, bug bounty programs, and coordinated disclosure efforts frequently lack sufficient detail to determine practical exploitability. 
Prior empirical studies of offensive security practices observe a shift away from purely black-box testing toward grey-box approaches that improve efficiency and coverage~\cite{happe2023understanding}. Incorporating a grey-box agentic exploitation framework such as AXE into detection pipelines allows maintainers to focus on issues for which exploitation is demonstrably feasible, supported by execution-based evidence available even in zero-day settings.

\textbf{Design Implications for Agentic Exploitation.}
The performance gap between AXE and the single-agent baseline indicates that separating high-level reasoning from source inspection and execution improves robustness in multi-step exploitation workflows. Failure attribution further shows that most unsuccessful runs fail at the planning stage rather than during execution. Tool augmentation helps compensate for some execution-level shortcomings by automating tasks such as endpoint discovery or payload construction, but it does not address the dominant reasoning failures observed in planning. These observations suggest that exploitation planning remains the most complex and error-prone component of agentic systems. Breaking planning into more specific and constrained subtasks may reduce this burden by forcing agents to reason over narrower hypotheses, explicitly enumerate feasible exploitation possibilities, and derive more targeted questions for exploration. Such decomposition also increases opportunities to replace parts of the workflow with more deterministic mechanisms, such as structured endpoint discovery or configuration analysis, which can further improve robustness. 
Recent work in vulnerability detection and exploitation has explored similar ideas by compelling language models to reason at finer granularity and guiding them 
with structured or neurosymbolic reasoning, pointing to a promising direction for future agentic exploitation systems~\cite{li2024iris, li2025towards, ling2025sound, liu2025llm}.

\textbf{Scope \& Generalizability.}
Our evaluation is conducted on CVE-Bench and therefore focuses on web applications with HTTP-based attack surfaces. However, the core evaluation model is not specific to the Web domain. The same detection-to-validation workflow can be applied to other classes of software. Under these conditions, AXE can be adapted without changing its core architecture, while swapping in interface-specific execution tooling and oracle definitions. While we expect similar gains in exploitation effectiveness in these environments, confirming this remains an important direction for future research.

\textbf{Environmental and Oracle-Specific Limitations.}
While AXE is designed for practical adoption, we acknowledge that automated validation requires system-specific setups. We confirm exploit success using reusable, predefined test cases that capture common impact patterns such as unauthorized file creation or data access. However, these oracles may not exhaustively cover all possible vulnerability impacts. For example, a path traversal vulnerability may permit file writes only under restricted names or locations, whereas our evaluation requires the agents to create a representative file to confirm impact. As a result, there may exist unique vulnerabilities whose effects are not captured by the available tests. Efficiency metrics are also limited to tools for which execution traces were available, so AvgTCA and efficiency are not reported for T-Agent and AutoGPT.

\textbf{Conservative Exploit Validation.} While AXE substantially improves over black-box approaches and yields perfect precision when exploit success is confirmed, it is not expected to identify all exploitable vulnerabilities. The approach is conservative by design, i.e., when a success condition is met, exploitability is guaranteed, but failure to confirm exploitation does not imply that a vulnerability is non-exploitable. As a result, some exploitable issues may remain unverified. Nevertheless, even identifying a small fraction of truly exploitable vulnerabilities from large volumes of unreviewed reports can provide significant practical value by directing developers' attention toward issues with confirmed security impact.

\textbf{Localization quality.}
Our results are contingent on the report providing a reasonably accurate suspected weakness category and source-code location. In the CVE-Bench evaluation, this context is manually constructed from public artifacts to model a localized report, so the results measure AXE's ability to confirm exploitability once such context is available. They do not measure the ability of upstream tools to produce accurate localization. The Letta case study provides initial evidence that AXE can also operate on noisier detector-produced locations, but evaluating localization sensitivity and broader detector-generated inputs remains future work.

\textbf{Ethical Considerations and Vulnerability Disclosure.}
All experiments were conducted in isolated, containerized environments and did not involve interaction with live or production systems. The benchmark evaluation uses previously disclosed vulnerabilities from CVE-Bench and does not introduce new disclosure obligations for those cases. The additional case study was conducted on a local instance of an open-source project and did not target any production deployment or real users. During the case study, AXE produced exploit evidence for a separate issue that, to our knowledge, did not have a public advisory or CVE at the time of analysis. The issue was present in the vulnerable version used for the case study, but the relevant code path had already been changed in the latest project version by replacing dynamic execution during schema derivation with static AST parsing. We contacted a project maintainer to ask about the change and have not received a response as of submission. In the submitted manuscript, we omit the full exploit payload and operational reproduction steps for this issue, and will provide additional disclosure information to the chairs if requested. AXE is intended to support defensive exploitability confirmation, vulnerability validation, and remediation prioritization.

\section{Related Work}
\label{sec:realted_word}

\subsection{LLM-based Exploit Automation}

Recent work has explored the use of LLMs as the decision-making component in automated exploitation and penetration testing. In non-web settings, LLM-driven agents have been shown to iteratively interact with real systems, generate commands, and adapt their behavior based on observed execution results in order to exploit vulnerabilities~\cite{happe2023getting, huang2023penheal, deng2023pentestgpt, fang2024one, happe2023llms, luong2025xoffense}. In the web domain, several systems frame exploitation as an interactive process in which agents probe endpoints, interpret responses, and attempt to exercise vulnerabilities in deployed applications rather than isolated code snippets~\cite{zhu2025cve, zhu2024teams, fang2024web}.

Recent systems have explored generating exploits directly from CVE information through a number of different approaches. FaultLine~\cite{nitin2025faultline} and CVE-GENIE~\cite{ullah2025cve} use CVE entries containing vulnerability descriptions and, in some cases, exploitation details as input to guide exploit generation, aiming to replicate reported vulnerabilities. Akhoundali et al.~\cite{akhoundali2025eradicating} have demonstrated end-to-end pipelines for detecting, exploiting, and remediating vulnerabilities, specifically for path traversal vulnerabilities, combining detection, exploitation, and downstream patching. POCGEN~\cite{simsek2025pocgen} shows that LLM-guided program analysis can generate executable PoC exploits for npm package vulnerabilities. PenForge~\cite{huang2026penforge} evaluates tool-augmented AutoGPT~\cite{richards2023autogpt} on the CVE-Bench dataset in a black-box setting, however, its evaluation differs from ours in few meaningful ways. PenForge uses a newer and more capable LLM model with post-disclosure training cutoffs, which introduces the problem of data contamination from previously disclosed vulnerabilities. It also permits external retrieval tools with internet access during exploitation and does not release full execution traces, complicating reproducibility and the interpretation of reported results.

\subsection{Agentic Framework Designs \& Failure Analysis}

Agentic LLM systems are commonly designed as an explicit interaction loop that alternates between reasoning about what to do next and taking tool-mediated actions to gather evidence or change the environment, instead of attempting to solve the task in a single prompt~\cite{masterman2024landscape, zhao2025llm, yao2022react}. ReAct is a representative pattern for this style of ``reason-and-act'' execution, where intermediate reasoning guides which actions to take and the resulting observations are fed back to update the next steps~\cite{yao2022react}. More broadly, surveys of agent architectures emphasize repeated design choices that support longer and more complicated tasks, including explicit planning and maintaining state over time~\cite{masterman2024landscape, zhao2025llm}. Many multi-agent systems further separate high-level decision making from lower-level tool execution through role specialization and a coordinating controller that delegates subtasks~\cite{masterman2024landscape, zhao2025llm}.

While prior work has examined failures in agentic systems mainly in the context of general-agent benchmarks and issue resolution~\cite{bouzenia2025understanding, cemri2025multi, zhang2025agent, liu2025process, liu2025empirical, sajadi2025safeaigeneratedpatcheslargescale, ehsani2025bugfixingbroadercontext}, vulnerability exploitation introduces distinct challenges that are not well represented in these settings. These challenges include identifying implicit preconditions, constructing vulnerability-specific payloads, and bridging high-level vulnerability metadata such as CWE labels to concrete, program-specific attack surfaces and exploitation paths. To our knowledge, this work is the first to systematically analyze agentic failures in the context of vulnerability exploitation. Additionally, prior work on failure attribution in agentic systems has proposed methods for identifying which agent induces failure and at what step within an execution trace~\cite{zhang2025agentracer}. In our study, we adapt these attribution techniques to the exploitation setting to analyze where and why multi-agent exploitation attempts break down.

\section{Conclusions}

This paper studies localized grey-box exploitability confirmation as a way to help developers triage and prioritize vulnerability reports. In this setting, a system receives minimal report context, such as a suspected weakness category and source-code location, along with source access and a running target, and attempts to produce executable evidence that the reported issue is exploitable. We instantiate this setting in AXE, which performs iterative planning, source inspection, sandboxed execution, and oracle-checked validation.

Our evaluation shows that localized grey-box report context substantially improves automated exploitability confirmation. AXE achieves 30\% Success@5 on CVE-Bench, compared with 10\% for black-box baselines, while role specialization further improves performance over a single-agent grey-box baseline under the same information boundary. For successful exploits, AXE generates PoC reports with reproducible steps and explicit verification oracles in 11 of 12 cases. Failure analysis shows that remaining failures are dominated by high-level reasoning gaps, especially misinterpreted vulnerability semantics, incorrect attack-surface selection, and unmet execution preconditions.

Overall, our results suggest that exploitability confirmation can bridge vulnerability reporting and remediation by helping developers prioritize reports with concrete exploit evidence. Future systems should focus less on low-level execution alone and more on semantic grounding, precondition discovery, and structured reasoning over source-level context.

\section*{Artifact Availability} 
To produce transparency and reproducibility, we have made the AXE implementation and all accompanying experimental artifacts publicly available at \url{https://anonymous.4open.science/r/310500it3r/README.md}. The release includes the source code and evaluation harness necessary to reproduce the results reported in this work. All experiments rely on publicly available datasets and containerized benchmarks.

\appendix
\section{Prompt Interfaces}
\label{app:prompts}

This appendix summarizes the prompt interfaces used by AXE. The complete prompt templates, prompt construction code, and schema definitions are included in the replication package. Here, placeholders written as \texttt{<field>} denote values dynamically inserted at runtime.

\begin{table*}[t]
\centering
\small
\caption{Prompt interface summary.}
\label{tab:prompt_interfaces}
\begin{tabularx}{\textwidth}{p{0.15\textwidth}X X p{0.22\textwidth}}
\toprule
\textbf{Component} & \textbf{System prompt context} & \textbf{User prompt context} & \textbf{Structured output} \\
\midrule
Strategist &
Role instructions for exploitation planning. In grey-box mode, the prompt includes the target URL, proof goals, safety constraints, and permission to request source-code exploration. In black-box mode, it removes source-code and CWE context and disallows exploration. &
Grey-box mode includes \texttt{<target>}, \texttt{<cwe>}, \texttt{<vuln\_summary>}, \texttt{<vuln\_file>}, \texttt{<vuln\_line>}, \texttt{<vulnerable\_file\_content>}, \texttt{<vulnerable\_snippet>}, \texttt{<exploration\_history>}, \texttt{<plan\_exploit\_history>}, \texttt{<endpoints>}, and \texttt{<budget\_status>}. Black-box mode includes prior plan and exploit history and known endpoints. &
\texttt{StrategyPlan} \\
\midrule
Explorer &
Role instructions for read-only repository inspection. The prompt includes \texttt{<source\_dir>}, the vulnerable location, and the vulnerable snippet. &
Includes \texttt{<commands\_so\_far>}, \texttt{<findings\_so\_far>}, \texttt{<commands\_left>}, \texttt{<questions\_text>}, \texttt{<components\_text>}, and \texttt{<endpoints>}. &
\texttt{ExplorerAction}, then \texttt{ExplorationSummary} \\
\midrule
Exploiter &
Role instructions for executing concrete commands against the target service. In grey-box mode, the prompt includes vulnerability context, source context, exploration results, exploitation strategy, target URLs, and proof goals. In black-box mode, it removes source-code and CWE context. &
Includes \texttt{<steps>}, prior command history \texttt{<hist>}, and environment values such as target URL, proof upload URL, secret file path, secret table name, default credentials, admin username, and outbound service URL. &
\texttt{ExploitStep}, then \texttt{ExploitationSummary} \\
\midrule
PoC Generator &
Role instructions for producing a concise Markdown proof-of-concept report from a successful exploitation trace. &
Includes \texttt{<application\_url>}, \texttt{<target\_info>}, \texttt{<cwe>}, \texttt{<vuln\_file>}, \texttt{<vuln\_line>}, \texttt{<vulnerable\_snippet>}, \texttt{<command\_trace>}, \texttt{<last\_command>}, and the relevant proof goals. &
\texttt{ExploitScript} \\
\midrule
Single-agent baseline &
Role instructions for a combined source-inspection and exploitation agent. The prompt includes source access, known endpoints, target metadata, vulnerability context, remembered findings, budget status, and proof goals. &
Includes \texttt{<steps>}, \texttt{<commands>}, truncated previous outputs \texttt{<hist>}, the full last output \texttt{<last\_full>}, discovered endpoints, and proof-related environment values. &
\texttt{SingleAgentStep}, then \texttt{ExploitScript} \\
\bottomrule
\end{tabularx}
\end{table*}

\noindent\textbf{Strategist output.}
The Strategist returns a plan and a decision indicating whether AXE should request more source-code context, execute an exploit attempt, or terminate.

\begin{lstlisting}[style=code,language=json]
{
  "actions": ["<exploitation step>"],
  "decision": "EXPLORE | EXECUTE | END",
  "thought": ["<planning rationale>"],
  "exploration_requests": [
    {
      "target_component": "<file, module, or service>",
      "question": "<source-level question>"
    }
  ]
}
\end{lstlisting}

\noindent\textbf{Explorer outputs.}
The Explorer either proposes one repository-inspection command or marks the exploration request as complete. After command execution, a separate summarization prompt condenses the observed evidence.

\begin{lstlisting}[style=code,language=json]
{
  "command": "<read-only shell command>",
  "thought": "<short rationale>",
  "done": true
}
\end{lstlisting}

\begin{lstlisting}[style=code,language=json]
{
  "summary": "<factual source-code findings>"
}
\end{lstlisting}

\noindent\textbf{Exploiter outputs.}
The Exploiter returns one executable command at a time. A null command indicates that it has no further command to run in the current attempt. After an attempt, AXE asks the Exploiter to summarize the observed command trace.

\begin{lstlisting}[style=code,language=json]
{
  "thought": "<short rationale>",
  "command": "<linux command or null>"
}
\end{lstlisting}

\begin{lstlisting}[style=code,language=json]
{
  "summary": "<attempt outcome and new findings>"
}
\end{lstlisting}

\noindent\textbf{Single-agent baseline output.}
The single-agent baseline uses the same source context, target context, tools, and proof goals as AXE, but combines planning, exploration, execution, and refinement in one prompt. Its output includes a short persistent finding field used as structured memory.

\begin{lstlisting}[style=code,language=json]
{
  "thought": "<short rationale>",
  "command": "<linux command or null>",
  "finding": "<short factual memory or null>"
}
\end{lstlisting}

\noindent\textbf{PoC report output.}
After a successful exploit, the PoC Generator produces a Markdown report from the observed successful trace. The report prompt asks for the affected component, trigger information, exploitation steps, verification oracle, evidence, and remediation guidance, without inventing unobserved URLs, parameters, or payloads.

\begin{lstlisting}[style=code,language=json]
{
  "script": "<Markdown proof-of-concept report>"
}
\end{lstlisting}

\bibliographystyle{IEEEtran}
\bibliography{ref}

\begin{thebibliography}{10}
\providecommand{\url}[1]{#1}
\csname url@samestyle\endcsname
\providecommand{\newblock}{\relax}
\providecommand{\bibinfo}[2]{#2}
\providecommand{\BIBentrySTDinterwordspacing}{\spaceskip=0pt\relax}
\providecommand{\BIBentryALTinterwordstretchfactor}{4}
\providecommand{\BIBentryALTinterwordspacing}{\spaceskip=\fontdimen2\font plus
\BIBentryALTinterwordstretchfactor\fontdimen3\font minus \fontdimen4\font\relax}
\providecommand{\BIBforeignlanguage}[2]{{%
\expandafter\ifx\csname l@#1\endcsname\relax
\typeout{** WARNING: IEEEtran.bst: No hyphenation pattern has been}%
\typeout{** loaded for the language `#1'. Using the pattern for}%
\typeout{** the default language instead.}%
\else
\language=\csname l@#1\endcsname
\fi
#2}}
\providecommand{\BIBdecl}{\relax}
\BIBdecl

\bibitem{medeiros2014automatic}
I.~Medeiros, N.~F. Neves, and M.~Correia, ``Automatic detection and correction of web application vulnerabilities using data mining to predict false positives,'' in \emph{Proceedings of the 23rd international conference on World wide web}, 2014, pp. 63--74.

\bibitem{guo2023mitigating}
Z.~Guo, T.~Tan, S.~Liu, X.~Liu, W.~Lai, Y.~Yang, Y.~Li, L.~Chen, W.~Dong, and Y.~Zhou, ``Mitigating false positive static analysis warnings: Progress, challenges, and opportunities,'' \emph{IEEE Transactions on Software Engineering}, vol.~49, no.~12, pp. 5154--5188, 2023.

\bibitem{dimastrogiovanni2016towards}
C.~Dimastrogiovanni and N.~Laranjeiro, ``Towards understanding the value of false positives in static code analysis,'' in \emph{2016 Seventh Latin-American Symposium on Dependable Computing (LADC)}.\hskip 1em plus 0.5em minus 0.4em\relax IEEE, 2016, pp. 119--122.

\bibitem{nadeem2012high}
M.~Nadeem, B.~J. Williams, and E.~B. Allen, ``High false positive detection of security vulnerabilities: a case study,'' in \emph{Proceedings of the 50th annual ACM Southeast Conference}, 2012, pp. 359--360.

\bibitem{zheng2025reviewers}
J.~Zheng, Y.~Zhou, A.~A. Ahmad, H.~Yao, and X.~Liu, ``From reviewers' lens: Understanding bug bounty report invalid reasons with llms,'' \emph{arXiv preprint arXiv:2511.18608}, 2025.

\bibitem{Abrams2026CurlAI}
L.~Abrams. (2026) Curl ending bug bounty program after flood of ai slop reports. https://www.bleepingcomputer.com/news/security/curl-ending-bug-bounty-program-after-flood-of-ai-slop-reports/.

\bibitem{Stenberg2025Slops}
D.~Stenberg. (2025) Death by a thousand slops. https://daniel.haxx.se/blog/2025/07/14/death-by-a-thousand-slops/.

\bibitem{laszka2016banishing}
A.~Laszka, M.~Zhao, and J.~Grossklags, ``Banishing misaligned incentives for validating reports in bug-bounty platforms,'' in \emph{European Symposium on Research in Computer Security}.\hskip 1em plus 0.5em minus 0.4em\relax Springer, 2016, pp. 161--178.

\bibitem{ruohonen2018bug}
J.~Ruohonen and L.~Allodi, ``A bug bounty perspective on the disclosure of web vulnerabilities,'' \emph{arXiv preprint arXiv:1805.09850}, 2018.

\bibitem{ideurope2025backlog}
{Information Development Europe B.V.}, ``Vulnerability management in crisis: Backlogs, zero-days, and global risks,'' 2025.

\bibitem{zhu2025cve}
Y.~Zhu, A.~Kellermann, D.~Bowman, P.~Li, A.~Gupta, A.~Danda, R.~Fang, C.~Jensen, E.~Ihli, J.~Benn \emph{et~al.}, ``Cve-bench: A benchmark for ai agents' ability to exploit real-world web application vulnerabilities,'' \emph{arXiv preprint arXiv:2503.17332}, 2025.

\bibitem{semgrep}
``{Semgrep: Static Analysis for Code},'' \url{https://semgrep.dev/}, Semgrep, Inc., 2026.

\bibitem{dirb}
``Dirb: Web content scanner,'' https://www.kali.org/tools/dirb/, OffSec Services Limited.

\bibitem{owasp_wstg_ptm_2025}
{OWASP Foundation}, ``Penetration testing methodologies,'' \url{https://owasp.org/www-project-web-security-testing-guide/latest/3-The_OWASP_Testing_Framework/1-Penetration_Testing_Methodologies}, OWASP Foundation, 2025.

\bibitem{happe2023understanding}
A.~Happe and J.~Cito, ``Understanding hackers’ work: An empirical study of offensive security practitioners,'' in \emph{Proceedings of the 31st ACM Joint European Software Engineering Conference and Symposium on the Foundations of Software Engineering}, 2023, pp. 1669--1680.

\bibitem{masterman2024landscape}
T.~Masterman, S.~Besen, M.~Sawtell, and A.~Chao, ``The landscape of emerging ai agent architectures for reasoning, planning, and tool calling: A survey,'' \emph{arXiv preprint arXiv:2404.11584}, 2024.

\bibitem{zhao2025llm}
B.~Zhao, L.~G. Foo, P.~Hu, C.~Theobalt, H.~Rahmani, and J.~Liu, ``Llm-based agentic reasoning frameworks: A survey from methods to scenarios,'' \emph{arXiv preprint arXiv:2508.17692}, 2025.

\bibitem{happe2023getting}
A.~Happe and J.~Cito, ``Getting pwn’d by ai: Penetration testing with large language models,'' in \emph{Proceedings of the 31st ACM joint european software engineering conference and symposium on the foundations of software engineering}, 2023, pp. 2082--2086.

\bibitem{yao2022react}
S.~Yao, J.~Zhao, D.~Yu, N.~Du, I.~Shafran, K.~R. Narasimhan, and Y.~Cao, ``React: Synergizing reasoning and acting in language models,'' in \emph{The eleventh international conference on learning representations}, 2022.

\bibitem{gao2024efficient}
S.~Gao, J.~Dwivedi-Yu, P.~Yu, X.~E. Tan, R.~Pasunuru, O.~Golovneva, K.~Sinha, A.~Celikyilmaz, A.~Bosselut, and T.~Wang, ``Efficient tool use with chain-of-abstraction reasoning,'' \emph{arXiv preprint arXiv:2401.17464}, 2024.

\bibitem{shi2024learning}
Z.~Shi, S.~Gao, X.~Chen, Y.~Feng, L.~Yan, H.~Shi, D.~Yin, P.~Ren, S.~Verberne, and Z.~Ren, ``Learning to use tools via cooperative and interactive agents,'' \emph{arXiv preprint arXiv:2403.03031}, 2024.

\bibitem{dang2025real}
W.~Dang, K.~Li, S.~Chen, Z.~Zhuo, L.~Zhang, and Z.~Liu, ``Real-world usability of vulnerability proof-of-concepts: A comprehensive study,'' \emph{arXiv preprint arXiv:2510.18448}, 2025.

\bibitem{wang2025aligning}
L.~Wang, W.~Dang, M.~Zhao, Y.~Wang, X.~Wu, and S.~Chen, ``Aligning core aspects: Improving vulnerability proof-of-concepts via cross-source insights,'' in \emph{Proceedings of the 33rd ACM International Conference on the Foundations of Software Engineering}, 2025, pp. 1774--1777.

\bibitem{nvd}
{National Institute of Standards and Technology}, ``National vulnerability database,'' \url{https://nvd.nist.gov}, 2025.

\bibitem{zhu2024teams}
Y.~Zhu, A.~Kellermann, A.~Gupta, P.~Li, R.~Fang, R.~Bindu, and D.~Kang, ``Teams of llm agents can exploit zero-day vulnerabilities,'' \emph{arXiv preprint arXiv:2406.01637}, 2024.

\bibitem{richards2023autogpt}
T.~B. Richards, ``Auto-gpt: An experimental open-source attempt to make gpt-4 fully autonomous,'' \url{https://github.com/Significant-Gravitas/Auto-GPT}, 2023.

\bibitem{owasp_zap}
{OWASP Foundation}, ``{OWASP Zed Attack Proxy (ZAP)},'' \url{https://www.zaproxy.org/}.

\bibitem{happe2023llms}
A.~Happe, A.~Kaplan, and J.~Cito, ``Llms as hackers: Autonomous linux privilege escalation attacks,'' \emph{arXiv preprint arXiv:2310.11409}, 2023.

\bibitem{el2025llm}
M.~A. El~Yagouby, A.~Lahmadi, M.~Zakroum, O.~Festor, and M.~Ghogho, ``Llm-cvx: A benchmarking framework for assessing the offensive potential of llms in exploiting cves,'' in \emph{Proceedings of the 18th ACM Workshop on Artificial Intelligence and Security}, 2025, pp. 194--205.

\bibitem{openai_gpt4o_docs_2026}
{OpenAI}, ``Gpt-4o model documentation,'' \url{https://platform.openai.com/docs/models/gpt-4o?utm_source=chatgpt.com}.

\bibitem{replication2026axe}
``Replication package,'' \url{https://anonymous.4open.science/r/310500it3r/}, 2026, anonymous 4open.science repository.

\bibitem{letta-ai_letta}
{letta-ai}, ``{Letta: Platform for Building Stateful AI Agents},'' \url{https://github.com/letta-ai/letta/}.

\bibitem{letta-static-parsing-fix}
\BIBentryALTinterwordspacing
S.~Wooders, ``fix: move to static parsing for python docstrings ({PR} \#3973),'' GitHub commit \texttt{136aa89}, Aug. 2025. [Online]. Available: \url{https://github.com/letta-ai/letta/commit/136aa890478e097539a25ae420286d56597aad87}
\BIBentrySTDinterwordspacing

\bibitem{wp_frontend_profile}
``Wp frontend profile,'' \url{https://wordpress.org/plugins/wp-front-end-profile/}.

\bibitem{li2024iris}
Z.~Li, S.~Dutta, and M.~Naik, ``Iris: Llm-assisted static analysis for detecting security vulnerabilities,'' \emph{arXiv preprint arXiv:2405.17238}, 2024.

\bibitem{li2025towards}
H.~Li, H.~Zhang, K.~Pei, and Z.~Qian, ``Towards more accurate static analysis for taint-style bug detection in linux kernel,'' in \emph{40th IEEE/ACM International Conference on Automated Software Engineering, ASE}, 2025.

\bibitem{ling2025sound}
Y.~Ling, G.~Rajiv, K.~Gopinathan, and I.~Sergey, ``Sound and efficient generation of $\{$Data-Oriented$\}$ exploits via programming language synthesis,'' in \emph{34th USENIX Security Symposium (USENIX Security 25)}, 2025, pp. 413--429.

\bibitem{liu2025llm}
P.~Liu, C.~Sun, Y.~Zheng, X.~Feng, C.~Qin, Y.~Wang, Z.~Xu, Z.~Li, P.~Di, Y.~Jiang \emph{et~al.}, ``Llm-powered static binary taint analysis,'' \emph{ACM Transactions on Software Engineering and Methodology}, vol.~34, no.~3, pp. 1--36, 2025.

\bibitem{huang2023penheal}
J.~Huang and Q.~Zhu, ``Penheal: A two-stage llm framework for automated pentesting and optimal remediation,'' in \emph{Proceedings of the workshop on autonomous cybersecurity}, 2023, pp. 11--22.

\bibitem{deng2023pentestgpt}
G.~Deng, Y.~Liu, V.~Mayoral-Vilches, P.~Liu, Y.~Li, Y.~Xu, T.~Zhang, Y.~Liu, M.~Pinzger, and S.~Rass, ``Pentestgpt: An llm-empowered automatic penetration testing tool,'' \emph{arXiv preprint arXiv:2308.06782}, 2023.

\bibitem{fang2024one}
R.~Fang, R.~Bindu, A.~Gupta, and D.~Kang, ``Llm agents can autonomously exploit one-day vulnerabilities,'' \emph{arXiv preprint arXiv:2404.08144}, 2024.

\bibitem{luong2025xoffense}
P.~D. Luong, L.~T.~G. Bao, N.~V.~K. Tam, D.~H.~N. Khoa, N.~H. Quyen, V.-H. Pham, and P.~T. Duy, ``xoffense: An ai-driven autonomous penetration testing framework with offensive knowledge-enhanced llms and multi agent systems,'' \emph{arXiv preprint arXiv:2509.13021}, 2025.

\bibitem{fang2024web}
R.~Fang, R.~Bindu, A.~Gupta, Q.~Zhan, and D.~Kang, ``Llm agents can autonomously hack websites,'' \emph{arXiv preprint arXiv:2402.06664}, 2024.

\bibitem{nitin2025faultline}
V.~Nitin, B.~Ray, and R.~Z. Moghaddam, ``Faultline: Automated proof-of-vulnerability generation using llm agents,'' \emph{arXiv preprint arXiv:2507.15241}, 2025.

\bibitem{ullah2025cve}
S.~Ullah, P.~Balasubramanian, W.~Guo, A.~Burnett, H.~Pearce, C.~Kruegel, G.~Vigna, and G.~Stringhini, ``From cve entries to verifiable exploits: An automated multi-agent framework for reproducing cves,'' \emph{arXiv preprint arXiv:2509.01835}, 2025.

\bibitem{akhoundali2025eradicating}
J.~Akhoundali, H.~Hamidi, K.~Rietveld, and O.~Gadyatskaya, ``Eradicating the unseen: Detecting, exploiting, and remediating a path traversal vulnerability across github,'' in \emph{Proceedings of the 20th ACM Asia Conference on Computer and Communications Security}, 2025, pp. 542--558.

\bibitem{simsek2025pocgen}
D.~Simsek, A.~Eghbali, and M.~Pradel, ``Pocgen: Generating proof-of-concept exploits for vulnerabilities in npm packages,'' \emph{arXiv preprint arXiv:2506.04962}, 2025.

\bibitem{huang2026penforge}
H.~Huang, J.~Shi, J.~Chen, T.~Zhang, Y.~Li, C.~Yang, E.~L. Ouh, L.~K. Shar, and D.~Lo, ``Penforge: On-the-fly expert agent construction for automated penetration testing,'' \emph{arXiv preprint arXiv:2601.06910}, 2026.

\bibitem{bouzenia2025understanding}
I.~Bouzenia and M.~Pradel, ``Understanding software engineering agents: A study of thought-action-result trajectories,'' \emph{arXiv preprint arXiv:2506.18824}, 2025.

\bibitem{cemri2025multi}
M.~Cemri, M.~Z. Pan, S.~Yang, L.~A. Agrawal, B.~Chopra, R.~Tiwari, K.~Keutzer, A.~Parameswaran, D.~Klein, K.~Ramchandran \emph{et~al.}, ``Why do multi-agent llm systems fail?'' \emph{arXiv preprint arXiv:2503.13657}, 2025.

\bibitem{zhang2025agent}
S.~Zhang, M.~Yin, J.~Zhang, J.~Liu, Z.~Han, J.~Zhang, B.~Li, C.~Wang, H.~Wang, Y.~Chen \emph{et~al.}, ``Which agent causes task failures and when? on automated failure attribution of llm multi-agent systems,'' \emph{arXiv preprint arXiv:2505.00212}, 2025.

\bibitem{liu2025process}
S.~Liu, Y.~Chen, R.~Krishna, S.~Sinha, J.~Ganhotra, and R.~Jabbarvand, ``Process-centric analysis of agentic software systems,'' \emph{arXiv preprint arXiv:2512.02393}, 2025.

\bibitem{liu2025empirical}
S.~Liu, F.~Liu, L.~Li, X.~Tan, Y.~Zhu, X.~Lian, and L.~Zhang, ``An empirical study on failures in automated issue solving,'' \emph{arXiv preprint arXiv:2509.13941}, 2025.

\bibitem{sajadi2025safeaigeneratedpatcheslargescale}
\BIBentryALTinterwordspacing
A.~Sajadi, K.~Damevski, and P.~Chatterjee, ``{How Safe Are AI-Generated Patches? A Large-scale Study on Security Risks in LLM and Agentic Automated Program Repair on SWE-bench},'' 2025. [Online]. Available: \url{https://arxiv.org/abs/2507.02976}
\BIBentrySTDinterwordspacing

\bibitem{ehsani2025bugfixingbroadercontext}
R.~Ehsani, E.~Parra, S.~Haiduc, and P.~Chatterjee, ``{Hierarchical Knowledge Injection for Improving LLM-based Program Repair},'' in \emph{40th IEEE/ACM International Conference on Automated Software Engineering (ASE)}, 2025.

\bibitem{zhang2025agentracer}
G.~Zhang, J.~Wang, J.~Chen, W.~Zhou, K.~Wang, and S.~Yan, ``Agentracer: Who is inducing failure in the llm agentic systems?'' \emph{arXiv preprint arXiv:2509.03312}, 2025.

\end{thebibliography}

\end{document}